\documentclass[nofootinbib,twocolumn]{an}
\usepackage{amsmath}
\usepackage{graphicx}
\usepackage{times}

\begin{document}

\title{Weak and strong field approximations and circular orbits of Kehagias-Sfetsos space-time}
\author{M. Dwornik \inst{1,2}\fnmsep\thanks{Corresponding author:
  \email{marek@titan.physx.u-szeged.hu}\newline} 
Zs. Horv\'{a}th\inst{1,2},
\and L.\'{A}. Gergely\inst{1,2,3} }

\authorrunning{M. Dwornik et al.}
\titlerunning{Weak and strong field approximations and circular orbits of Kehagias-Sfetsos space-time}

\institute{
Department of Theoretical Physics, University of Szeged, Tisza Lajos krt
84-86, Szeged 6720, Hungary
\and
Department of Experimental Physics, University of Szeged, D\'{o}m t\'{e}r 9,
Szeged 6720, Hungary
\and
Department of Physics, Tokyo University of Science, Shinjuku-ku, Tokyo, Japan
}


\abstract{The Kehagias-Sfetsos asymptotically flat black hole and naked singularity solutions of Ho\v{r}ava-Lifshitz 
gravity are investigated both in the weak-field and strong-field regimes. In the weak-field limit the gravitational field 
generated by the Kehagias-Sfetsos spherically symmetric solution is weaker then in the case of the Schwarzschild 
black hole of general relativity. In the strong-field regime naked singularities with $\omega_{0} \ll 1 $ display an 
unusual distance dependence: gravity becomes weaker when approaching the singularity.
The stability of circular orbits is also analyzed. While in the black hole case the square of the angular momentum 
should be larger than a certain finite, non-zero minimal value, in the naked singularity case there are stable 
circular orbits for any non-zero value of the angular momentum. 
In this regime we prove the existence of an infimum of the allowed radii of circular orbits 
(corresponding to vanishing angular momentum).}

\maketitle

\section{Introduction}

General relativity (GR) has been precisely tested on the Solar system scale,
however the very small and very large distance behaviour of gravity is less
well verified, leading to numerous proposed modifications of GR. Recently Ho\v{r}ava 
proposed a modification of GR at high energies, motivated by the
Lifshitz scalar field theory in solid state physics. The Ho\v{r}ava-Lifshitz
(HL) gravitational theory introduces aniso\-tropy between space and time. A
recent review of its Lorentz invariance violation, occurring at
trans-Planckian energy \\scales is presented in (\cite{visser2011}). Among the
several proposed versions of the HL theory, the infrared (IR)-modified Ho\v{r}ava 
gravity is the one which seems to be consistent with the current
observational data (\cite{kono2009,chen2009,chenwang}).

The spherically symmetric space-time in vacuum HL gravity is characterized
by the family of metrics (\cite{radi}) 
\begin{equation}
ds^{2}=-f(r)dt^{2}+f^{-1}(r)dr^{2}+r^{2}(d\theta ^{2}+\sin ^{2}\theta
d\varphi ^{2})\;,
\end{equation}
with 
\begin{equation}
f(r)=1+(\omega -\Lambda )r^{2}-\sqrt{r\left[ \omega \left( \omega -2\Lambda
\right) r^{3}+\beta \right] }\;.  \label{metric}
\end{equation}
Here $\beta $ is an integration constant, while $\Lambda $ and $\omega $ are
real parameters. Depending on the values of $\beta $ , $\omega $ and 
$\Lambda $, special cases of the metric (\ref{metric}) arise.

The choice $\beta =-(\alpha ^{2})/(\Lambda )$ and $\omega =0$ leads to the
Lu-Mei-Pope black hole solution (\cite{lu}). With $\beta =4\omega m$ and 
$\Lambda =0$ the Kehagias-Sfetsos (KS) solution is obtained (\cite{keha2009}). 
This paper focuses on the latter space-time, for which the function $f(r)$
is 
\begin{equation}
f(r)=1+\omega r^{2}-\sqrt{\omega ^{2}r^{4}+4\omega mr}\;.  \label{keha}
\end{equation}
This solution is asym\-ptotically flat and for large $r$ or when $\omega
\rightarrow \infty $ it approaches the Schwarzschild solution of GR. A
slowly rotating modification of the Kehagias-Sfetsos solution was introduced
in \cite{leekim}.

Beyond mass $m$, the space-time (\ref{keha}) is characterised by the 
Ho\v{r}ava-Lifshitz parameter $\omega >0$. It is customary to employ the
dimensionless parameter $\omega _{0}=\omega m^{2}>0$ too. General relativity
is recovered for $\omega _{0}$ $\rightarrow \infty $ but the black hole
interpretation continues to hold for any $\omega _{0}\geq 0.5$. Indeed, when 
$\omega _{0}>1/2$ there are two event horizons at 
\begin{equation}
r_{\pm }=m\left( 1\pm \sqrt{1-\frac{1}{2\omega _{0}}}\right) \;.
\end{equation}
The two event horizons coincide for $\omega _{0}=0.5$. The space-time 
(\ref{keha}) becomes a naked singularity whenever $\omega _{0}<0.5.$

The value of the parameter $\omega _{0}$ has been constrained by various
methods. The radar echo delay in the Solar system, analyzed in 
(\cite{harko2011}) gave the limit $\omega _{0,\min }^{(red)}=2\times 10^{-15}$.
The analysis of the perihelion precession of Mercury and of the deflection
of light by the Sun resulted in $\omega _{0,\min }^{(pp)}=6.9\times 10^{-16}$
and $\omega _{0,\min }^{(ld)}=1.1\times 10^{-15}$, respectively. Tighter
constraints for $\omega _{0}$ were presented in (\cite{iorio2010}), based on
the analysis of the range-residuals of the planet Mercury: $\omega _{0,\min
}^{(residual)}=7.2\times 10^{-10}$. A slightly stronger constraint $\omega
_{0,\min }^{(Sag)}=8\times 10^{-10}$ arises from the observation of the S2
star orbiting the Supermassive Black Hole (Sagittarius A*) in the center of
our Galaxy. It has been shown in (\cite{horvath2011}) that the forthcoming
Large Synoptic Survey Telescope will be able to constrain $\omega _{0}$ up
to $10^{-1}$ from strong gravitational lensing. We remark that neither of
these observations could render the limiting value of $\omega _{0}$ into the
regime where the Kehagias-Sfetsos solution describes a black hole.

Charged particles in orbital motion about a compact astrophysical bo\-dy
form an accretion disk. The simplest accretion disk model is the
steady-state thin disk, based on several simplifying assumptions 
(\cite{chen2012}). For accretion disks with sub-Eddington luminosities the inner
edge of the accretion disk is located at the innermost stable circular orbit
(ISCO) (\cite{pac2000}). However, for larger accretion disk luminosities,
there is no uniquely defined inner edge. Different definitions lead to
different edges, the differences increasing with the luminosity. In 
(\cite{abra2010}) six possible definitions of the inner edge have been listed.

Accretion characteristics in the IR limit of HL gravity, based on the thin
disk model were explored both for spherically symmetric (\cite{harko2009})
and for slowly rotating (\cite{harko2011}) black holes. In the spherically
symmetric case the energy flux, the temperature distribution of the disk and
the spectrum of the emitted black body radiation all significantly differ
from the GR predictions. Also, the intensity of the flux emerging from the
disk surface is larger for the slowly rotating KS solution than for the
rotating Kerr black hole of GR.

This paper revisits various aspects related to accretion in a KS space-time
and is organized as follows. In Sections 2 and 3 we investigate the KS
metric in the weak-field and strong-field limits, respectively. In Section 4
we analyze the properties of the stable circular orbits in both the black
hole and the naked singularity parameter regimes. We summarize our results
in Section 5.

\section{Weak-field regime: weakened gravity}

The smallness of the post-Newtonian parameter $\varepsilon =m/r$ (in units $G=1=c$) 
characterizes the weak-field regime. Even with $\varepsilon \ll 1$,
the relative magnitude of the two parameters $\varepsilon $ and $\omega _{0}$
leads to three different limits.

When $\omega _{0}\gg \varepsilon ^{3}$ then $\omega _{0}^{-1}\varepsilon
^{3}\ll 1$ and the Kehagias-Sfetsos metric function can be rewritten as 
\begin{eqnarray}
f &=&1+\omega _{0}\left( \frac{r}{m}\right) ^{2}\left[ 1-\left( 1+\frac{
4m^{3}}{\omega _{0}r^{3}}\right) ^{1/2}\right]  \notag \\
&=&1+\omega _{0}\varepsilon ^{-2}\left[ 1-\left( 1+\frac{4}{\omega _{0}}
\varepsilon ^{3}\right) ^{1/2}\right] \;.  \label{1}
\end{eqnarray}
Expanding the expression in the bracket as 
\begin{equation}
\left( 1+\frac{4}{\omega _{0}}\varepsilon ^{3}\right) ^{1/2}\approx 1+\frac{2}{\omega _{0}}\varepsilon ^{3}\;,
\end{equation}
the metric function becomes
\begin{equation}
f\simeq 1-2\varepsilon \;,
\end{equation}
thus in this parameter regime the KS\ metric approximates the Schwarzschild
metric.

\begin{figure}[t]
\includegraphics[height=4.8cm, angle=360]{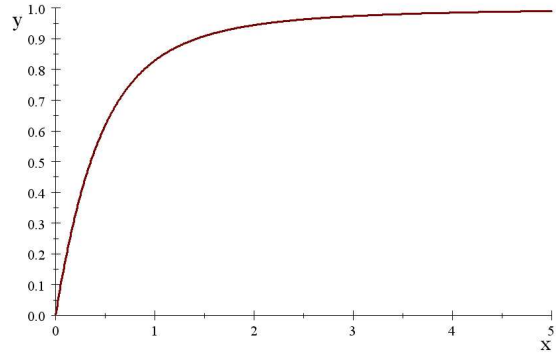}
\caption{The effective mass parameter in the $\protect\varepsilon \approx 
\protect\omega _{0}^{1/3}\ll 1$ regime as function of $x=\left( \protect
\omega _{0}\protect\varepsilon ^{-3}\right) ^{1/2}/2$.}
\label{y}
\end{figure}

When $\omega _{0}\approx \varepsilon ^{3}$ then $\omega _{0}^{-1}\varepsilon
^{3}=\mathcal{O(}1\mathcal{)}$, thus the Taylor series expansion cannot be
performed as in the previous case. Nevertheless $\omega _{0}\varepsilon
^{-2}=$ $\omega _{0}\varepsilon ^{-3}\varepsilon \simeq \varepsilon $ and
the metric function can be rewritten as%
\begin{eqnarray}
f &=&1+\varepsilon \left\{ \omega _{0}\varepsilon ^{-3}-\left[ \left( \omega
_{0}\varepsilon ^{-3}\right) ^{2}+4\omega _{0}\varepsilon ^{-3}\right]
^{1/2}\right\}  \notag \\
&=&1-2\varepsilon \left\{ \left[ \omega _{0}\varepsilon ^{-3}+\left( \frac{
\omega _{0}\varepsilon ^{-3}}{2}\right) ^{2}\right] ^{1/2}-\frac{\omega
_{0}\varepsilon ^{-3}}{2}\right\}  \notag \\
&=&1-2\varepsilon \mathcal{O}(1)\;.
\end{eqnarray}
The curly bracket in the second line (to be denoted as $y$) can be rewritten
in terms of $x=\left( \omega _{0}\varepsilon ^{-3}\right) ^{1/2}/2$ as 
\begin{equation}
y=2x\left[ \left( 1+x^{2}\right) ^{1/2}-x\right]
\end{equation}
and shown on Fig. \ref{y} to take values in the interval $\left( 0,1\right) $
only. Therefore the weak-field approximaton of the KS solution in this
parameter range is again the Schwarzschild solution, nevertheless with an
effective mass parameter which is smaller than the mass. 

Finally when $\omega _{0}\ll \varepsilon ^{3}$ then $\omega
_{0}^{-1}\varepsilon ^{3}\gg 1$, and we get by another series expansion 
\begin{eqnarray}
f &=&1+\left( \omega _{0}\varepsilon ^{-3}\right) \varepsilon \left[
1-\left( 1+\frac{4}{\omega _{0}}\varepsilon ^{3}\right) ^{1/2}\right]  \notag
\\
&\simeq &1-2\varepsilon \left( \omega _{0}\varepsilon ^{-3}\right) ^{1/2}\;.
\end{eqnarray}
This is another Schwarzschild regime with a very small effective mass
parameter.

In summary in the weak field limit the metric function becomes 
\begin{equation}
f=1-2\varepsilon y\;,
\end{equation}
with the positive parameter 
\begin{eqnarray}
y &=&1,~\text{for}~\omega _{0}\gg \varepsilon ^{3}~,  \notag \\
y &<&1,~\text{for}~\omega _{0}\approx \varepsilon ^{3}~,  \notag \\
y &\ll &1,~\text{for}~\omega _{0}\ll \varepsilon ^{3}~.
\end{eqnarray}
In spite of the fact that for all $\omega _{0}<0.5$ the KS\ space-time
represents a naked singularity, its weak-field regime is Schwarzschild with
a positive effective mass 
\begin{equation}
m_{eff}=my\;.
\end{equation}
As $m_{eff}\leq m$, gravity is weaker in the weak-field limit of the KS
space-time compared to the Schwarzschild case$.$

\section{Strong-field regi\-me: unusual distance dependence}

Approaching the center the value of the post-Newtonian parameter increases,
at the Schwarzschild radius becoming \thinspace $0.5$. Hence in this
strong-field regime $\varepsilon =\mathcal{O(}1\mathcal{)}$. When $\omega
_{0}\gg 1$, a series expansion of Eq. (\ref{1}) in $\omega _{0}^{-1}$ leads
to the Schwarzschild metric function $f=1-2\varepsilon $. The regime $\omega
_{0}=\mathcal{O(}1\mathcal{)}$ has been studied numerically (\cite{harko2009}
) for the black hole case $\omega _{0}>0.5$.

\begin{figure}[t]
\includegraphics[height=6.8cm, angle=360]{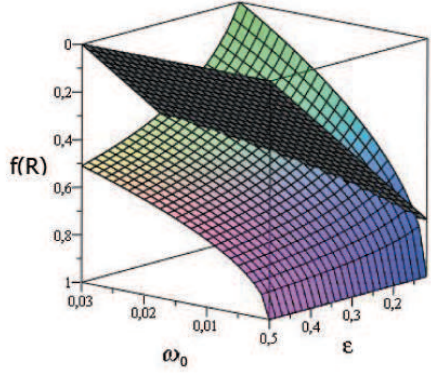}
\caption{The behaviour of the metric function $f$ (lighter surface) as
compared to the respecting metric function of the Schwarzschild black hole
(dark surface). For large values of $\protect\varepsilon $ and small values
of $\protect\omega _{0}$ the $1/r$ dependence (linear dependence on $\protect
\varepsilon $)\ of $f$ of the Schwarzschild space-time is replaced by an $
r^{1/2}$ dependence ($\protect\varepsilon ^{-1/2}$ dependence) in the KS
space-time.}
\label{squarer}
\end{figure}

When $\omega _{0}\ll 1$ (which is in the naked singularity regi\-me) a
series expansion of Eq. (\ref{1}) in $\omega _{0}^{-1}$ leads to 
\begin{equation}
f\simeq 1-2\varepsilon ^{-1/2}\omega _{0}^{1/2}=1-2\left( \frac{\omega _{0}}{
m}\right) ^{1/2}r^{1/2}\;.  \label{gttstrong1}
\end{equation}
(Note that $f$ stays positive as $\varepsilon >4\omega _{0}$ is obeyed in
the chosen parameter range.) Hence gravity decreases when approaching the
singularity. This behaviour is radically different from the GR prediction
and is presented in Fig.~\ref{squarer}.

\section{Circular orbits about Kehagias-Sfetsos black holes}

The observational tests on the limiting $\omega _{0}\approx 10^{-\left(
9\div 16\right) }$ confirm that in the $\omega _{0}=$ $\mathcal{O(}1\mathcal{
)}$ regime the geodetic motion of free particles is a valid approximation (
\cite{mosaffa2011}). In this section we focus to this parameter range,
studying the timelike geodetic motion in the KS space-time and determining
the radius of the innermost stable circular orbit (the inner edge of the
accretion disk).

Circular motions in the equatorial plane $(\theta =\pi /2)$ are defined by a
constant normalized radial coordinate $R:=r/m$. The $t,$ $r$ and $\varphi $
components of the geode\-sic equation reduce to 
\begin{equation}
\left( \frac{d^{2}\varphi (\tau )}{d\tau ^{2}}\right) =0,  \label{geod1}
\end{equation}
\begin{equation}
\left( \frac{d^{2}t(\tau )}{d\tau ^{2}}\right) =0,  \label{geod2}
\end{equation}
\begin{equation}
\frac{\omega _{0}\left( R\Xi -\omega _{0}R^{3}-1\right) }{m^{2}\Xi }\left( 
\frac{dt(\tau )}{d\tau }\right) ^{2}=R\left( \frac{d\varphi (\tau )}{d\tau }
\right) ^{2},  \label{geod3} 
\end{equation}
respectively, where $\Xi =\sqrt{R\omega _{0}\left( R^{3}\omega _{0}+4\right) 
}$ and $\tau $ is the proper time. Eqs. (\ref{geod1}) and (\ref{geod2})
imply that $\dot{\varphi}(\tau )= \\d\varphi (\tau )/d\tau $ and $\dot{t}(\tau
)=dt(\tau )/d\tau $ are constants, leading to conserved specific orbital
angular momentum \\$L=m^{2}R^{2}\dot{\varphi}$ and specific energy $E=f\dot{t}$ of
the particle. Eq. (\ref{geod3}), through the positivity of $\omega _{0}$
implies 
\begin{equation}
R>\left( \frac{1}{2\omega _{0}}\right) ^{1/3}~.  \label{ineq}
\end{equation}
Both for stable or unstable circular orbits, regardless \\whether the KS
space-time represents a black hole or a na\-ked singularity, Eq. (\ref{ineq})
holds as a strict inequality. Note that the equality would imply $L=0$,
while smaller values of $R$ would lead to the forbidden range $L^{2}<0$.

The effective potential can be expressed as \cite{harko2009}:
\begin{equation*}
V_{eff}(L,R,\omega _{0})=
\end{equation*}
\begin{equation}
\left[ 1+\omega _{0}R^{2}\left( 1-\sqrt{1+\frac{4}{\omega _{0}R^{3}}}\right) 
\right] \left( 1+\frac{L^{2}}{R^{2}}\right) ~,  \label{veff}
\end{equation}
a constant of motion itself ($dV_{eff}/dR=0$), as it depends only on $L,~R$
and $\omega _{0}$. Stable circular orbits occur at the local minima of the
potential, while the local maxima in the potential are the locations of
unstable circular orbits.

\subsection{Black hole parameter range ($\protect\omega _{0}\geq 0.5$)}

\begin{figure}[t]
\begin{center}
\includegraphics[height=7cm, angle=270]{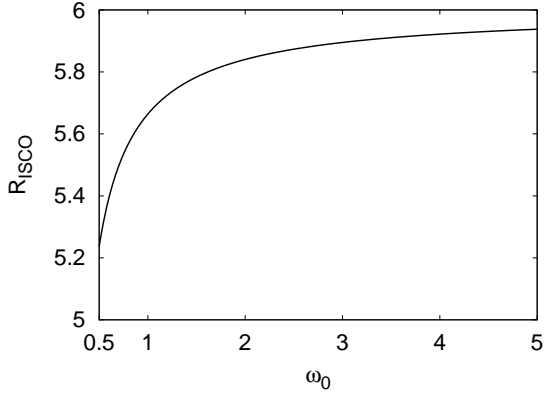}
\end{center}
\caption{The ISCO radius as a function of the dimensionless parameter $
\protect\omega _{0}$ in the black hole regime. For large $\protect\omega 
_{0} $ values the radius of the ISCO approaches $R_{ISCO}=6$, characteristic
for the Schwarzschild black hole.}
\label{iscobh}
\end{figure}

For each $\omega _{0}\geq 0.5$ a value of the angular momentum $L$ can be
found for which there is one extremum of the effective potential. At this
radius the conditions $dV_{eff}/dR=0$ and $d^{2}V_{eff}/dR^{2}=0$ are
obeyed. These equations can be solved numerically as follows. From the
extremum condition $dV_{eff}/dR=0$ one can express $L=L(R,\omega _{0})$,
which when inserted in the marginal stability condition \newline
$d^{2}V_{eff}/dR^{2}=0$ gives an implicit relation $F\left( R,\omega
_{0}\right) =0$, numerically solved as $R=R(\omega _{0})$. This mar\-ginally
stable orbit is the ISCO, the radius of which is represented as function of
the parameter $\omega _{0}$ on Fig~\ref{iscobh}.

\subsection{Naked singularity parameter range ($\protect\omega _{0}<0.5$)}

\begin{figure}[t]
\begin{center}
\includegraphics[height=7cm, angle=270]{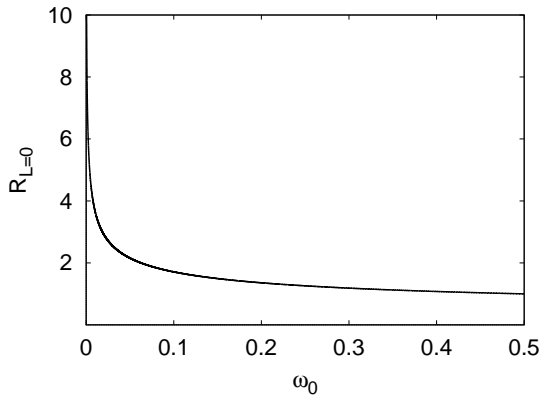}
\end{center}
\caption{The $\left( 2\protect\omega _{0}\right) ^{-1/3}$ curve, where $%
L^{2}=0$. Stable circular orbits exist for radii lying above this curve. }
\label{iscons}
\end{figure}

The numerical study shows that stable circular orbits can exist for any
non-zero angular momentum, as long as the inequality (\ref{ineq})\ holds.
This means that the set of the allowed stable circular orbit radii have an
infimum (a value which can be approached arbitrarily close, but cannot be
reached exactly) at $\left( 2\omega _{0}\right) ^{-1/3}$.\ The location of
the infimum as function of $\omega _{0}$ is shown on Fig~\ref{iscons} .

We note here that the ISCO radius represented on Fig. 4 of Ref. \cite%
{aah2011} seems to run below our curve, in the forbidden range $L^{2}<0$.

\section{Conclusions}

We investigated the Kehagias-Sfetsos asymptotically flat \\and spherically
symmetric solution of Ho\v{r}ava-Lifshitz gravity both in the weak and
strong-field regimes. This solution reduces to the Schwarzschild solution of
general relativity in the $\omega _{0}\rightarrow \infty $ limit of the Ho%
\v{r} ava-Lifshitz parameter $\omega _{0\text{.}}$ For $\omega _{0}\geq 0.5$
it represents a black hole, while for $0<\omega _{0}<0.5$ it is a naked
singularity.

We analyzed the weak-field regime (characterized by the smallness of the
post-Newtonian parameter $\varepsilon =m/r$) for generic $\omega _{0}$,
finding that gravity is weaker in the \\Kehagias-Sfetsos space-time than in
the Schwarzschild space-time. Then we have shown that in the strong-field
regime (close to the central singularity, where $\varepsilon =\mathcal{O(}1
\mathcal{)}$) of Kehagias-Sfetsos naked singularities with $\omega _{0}\ll 1$
gravity surprisingly decreases as $r^{1/2}$ while approaching the center.

We also studied the timelike geodesics in the black hole and naked
singularity regimes. In the black hole parameter range the ISCO radius as
function of $\omega _{0}$ has been studied in Ref. \cite{harko2009}. We have
extended the discussion into the naked singularity parameter range, finding
that stable circular orbits always exist if the angular momentum is
non-vanishing, implying that only an infimum of the stable circular orbit
radii can be defined whenever $\omega _{0}<0.5$.

Based on these results, in a subsequent work we will investigate the energy
flux of the accretion disk in the naked singularity case.

\acknowledgements This work was partially supported by the European Union
and European Social Fund grants T\'{A}MOP-4.2.2.A-11/1/KONV-2012-0060 and 
T\'{A}MOP-4.2.2/B-10/1-2010-0012. \\L\'{A}G was supported by EU grant T\'{A}MOP-4.2.2.A-11/1/KONV-2012-0060 and the Japan Society for the Promotion of Science.

\end{document}